\documentclass{ws-procs9x6}
\usepackage{epsfig} 
\newcommand{\quq}{\theta_{13}}
\newcommand{\nmt}{\nu_\mu \rightarrow \nu_\tau}
\newcommand{\nme}{\nu_\mu \rightarrow \nu_e}

\newcommand{\dmsq}{\Delta m^2}

\begin{document}

\title{Plans for Experiments to Measure $\theta_{13}$}

\author{Maury Goodman}

\address{Argonne National Laboratory\\Argonne IL 60439, USA\\
E-mail: maury.goodman@anl.gov}

\maketitle

\abstracts{New experiments at accelerators
and reactors are being designed to search for a
possible non-zero value of the MNS matrix parameter $\theta_{13}$.}

\section{Neutrino Parameters and the MNS matrix}
The neutrino mixing matrix, analogous to the CKM matrix for quarks, is
the MNS matrix\cite{bib:mns}.  In the simplest case, that matrix can
be parameterized by three angles and one CP violating phase.  With the
remarkable recent progress in understanding neutrinos, it
is known that at least two of these mixing angles are large, unlike
the angles in the quark sector.  Despite these large angles,
neutrino oscillation analyses are usually
based on the assumption that there are only two neutrinos.  This
simplifies the analyses, and is accurate, but can be confusing.
\par The three mixing angles are designated
$\theta_{12}$, $\theta_{23}$ and $\theta_{13}$.  Solar neutrino experiments
and the KamLAND reactor experiment have measured $\theta_{12}$, 
also called $\theta_{sol}$.  Atmospheric neutrino experiments have
been used to measure $\theta_{23}$ or $\theta_{atm}$.
Long-baseline accelerator neutrino oscillation experiments have been designed
to measure this angle.  The third mixing angle, $\theta_{13}$, has only 
been limited to be smaller than $\sin^2(2\theta_{13})~<~0.17$ by the
CHOOZ\cite{bib:chooz} nuclear reactor experiment, at the currently favored value of
$\Delta m^2_{atm}$.     A CP violating phase which appears in the MNS matrix
is called $\delta$.  Long-baseline neutrino experiments would be
sensitive to a non-zero value of $\theta_{13}$ by measuring $\nme$
oscillations with a rate smaller than the dominant expected $\nmt$
oscillations, so this is often called a sub-dominant process.

\par The current program of long-baseline neutrino experiments includes the
K2K experiment in Japan, which has been running since 1998, the NuMI/MINOS
program at Fermilab, which will start in 2005, and the CNGS program
at CERN which is expected to start in 2006.  Accelerator beams consist mostly of muon 
neutrinos ($\nu_\mu$) which are made when pions and Kaons decay.  
All three experiments expect to measure a change in the number and
distribution of $\nu_\mu$ as a result of $\nmt$ oscillation.   In addition,
CNGS hopes to measure $\nu_\tau$ appearance.  These $\nu_\tau$s would be present
in all three experiments, but the production of $\tau$s in K2K and MINOS
is expected to be suppressed by kinematic thresholds.  

\section{The importance of $\quq$}
\par I share the view that
the single most important task before the neutrino 
community is to measure a non-zero value for $\theta_{13}$ if
it exists.  While
$\sin^2(2 \theta_{23})$ is near 1, SNO data supports that
$\sin^2(2 \theta_{12})$ is approximately $0.7$ and is clearly not maximal.
Although we don't know what symmetries if any control the values of
the MNS mixing angles, it
is reasonable to expect that if $\theta_{12}$ is not maximal,
$\theta_{13}$ is probably not exactly zero.  The current limit on
$\sin^2(2 \theta_{13})$ is $<~0.17$\cite{bib:chooz}.  Some of us 
suspect that a new search
which is sensitive down to 0.02 would have a 90\% chance of
measuring a non-zero value for $\theta_{13}$!
\par Such a result would open up new possibilities for future neutrino
research.
If $\theta_{13}$ is not zero, searches for matter effects and
CP violation with accelerator neutrinos are feasible, in particular
using $\nme$ measurements as described below in Equation \ref{eq:p}.
The observation of matter effects in
$\nu$ or $\bar{\nu}$ tells us if $\Delta m^2_{31}$ is positive
(normal mass hierarchy) or negative (inverted mass hierarchy).  Of
even greater interest, the measurement of differences between
$\nu$ and $\bar{\nu}$ oscillation parameters (after removing
any matter effects) is sensitive to CP violating effects in the
lepton sector.  The measure of CP violation is the
Jarlskog invariant, which is proportional to factors which include
the product of all
mixing angles and $\Delta m^2$ values.  With the large mixing in the
lepton sector, CP violation in the neutrino sector is poised to be
50 times larger than in the quark sector, depending solely on whether
$\quq$ is near is current upper limit, or is much smaller.

\par The probability in vacuum for $\nme$ is\cite{bib:barger}:
\begin{eqnarray}
P(\nme) & = & 2 \sin(2\quq)s_{23}c_{13}s_{12}(s_{12}s_{23}s_{13}
-c_{12}c_{23}c_\delta) \sin^2\phi_{32} \\
& + & 2\sin(2\quq) s_{23}c_{13}c_{12}(c_{12}s_{23}s_{13}
+ s_{12}c_{23}c_\delta)\sin^2\phi_{31} \nonumber \\
& - & 2\sin(2\theta_{12})c^2_{13}[s_{12}c_{12}
(s^2_{13}s^2_{23}-c^2_{23}) + s_{13}s_{23}c_{23}(s^2_{12}-c^2_{12})
c_\delta]\sin^2\phi_{21} \nonumber \\
& + & \frac{1}{2} \sin(2\theta_{12})\sin(2\quq)\sin(2
\theta_{23})c_{13}s_\delta[\sin\phi_{32}\cos\phi_{32} \nonumber \\
& - & \sin\phi_{31}\cos\phi_{31}+\sin \phi_{21} \cos\phi_{21}] \nonumber
\end{eqnarray} 
 where $\phi_{ij} = \Delta m^2_{ij}L/(4E)$ and $c_{ij}$ and 
$s_{ij}$ refer
to the cosine and sine of the mixing angle ij.  The first two terms
describe the behavior at an L/E corresponding to the large value
of $\Delta m^2$ and the terms proportional to $\sin\delta$ ($\cos\delta$)
are CP odd (even).
\par In contrast reactor neutrino disappearance at small distances
(i.e. L/E $\sim \Delta m^2_{31}$, so L $\sim$ 1 km), the formula is
the simpler familiar one:
\begin{equation}
P(\bar{\nu}_e \rightarrow \bar{\nu}_e) \cong 
1 - sin^2(2\quq) sin^2 \phi_{32}
\end{equation}

\section{Future Off-Axis Long-Baseline Neutrino Experiments}
The search for a few percent $\nme$ in long-baseline experiments is made difficult
by three backgrounds.   First, there are $\nu_e$ in the beam near
the percent level from K and $\mu$ decays in the beam pipe.
Second, when the dominant $\nmt$ oscillation takes place and a $\tau$
is produced by charged current interaction, the $\tau$ decays to an
electron 17\% of the time.  Third, for any detector which resembles
a calorimeter, some fraction of the neutral current events are going to
be indistinguishable from an electron, particularly the single $\pi^0$
neutral current events.  
\par The kinematics of neutrino production is such that off the main axis
of the neutrino beam, the average neutrino energy is both lower and more
peaked toward a single value.  The lower energy means that event rates are
way down, already a significant limitation
 for long-baseline neutrino experiments.
However, an off-axis narrow-band beam can 
be used to reduce all three backgrounds.
The $\nu_e$ in the beam are mostly at high energy, and do not constitute a
large background to the expected signal.  There are no $\nu_\tau$ charged
current interactions below 4 GeV, so the tau decay background is greatly
reduced, and though the neutral current background is not eliminated, it
is greatly reduced since there is no high energy tail on the neutrino
spectrum.  This has led to consideration of new accelerator
based neutrino programs in the US and Japan.
\subsection{T2K}
The Japanese Particle Research Center (JPARC) is a new 50 GeV
proton synchrotron being built in Tokai.  A neutrino
beam is being planned.\cite{bib:t2k}
  In a first phase the accelerator would operate
at 0.77 MW, but an upgrade to 4 MW 
is being considered.  The 22.5 kiloton Super-Kamiokande detector
is 295 km away, and the beam could be built to be simultaneously
a few degrees off axis to that experiment and to the proposed site
for a new 1000 kiloton Hyper-Kamiokande detector in Tochibora.
With a 5 year run of JPARC, and the proposed 2 degree off-axis
beam, T2K would be able to measure $\quq$ or set a limit
on $\sin^2(2\theta_{13})~<~0.006$ at 90\%~CL for $\delta = 0$.  
The Hyper-Kamiokande detector
would be similar in design to Super-K, using large 50 cm diameter
Hamamatsu phototubes.
\subsection{NO$\nu$A}
\par A proposal is being developed for an off-axis experiment using
the NuMI beam at Fermilab.\cite{bib:p929}  Any detector would be built near
the surface of the earth, about 10 km away from the center of the NuMI
beam.  The detector would have
a mass of 50 Kilotons, and be sensitive to 1 GeV electron showers.
The passive detector is planned to be 7 sheets of 2.5 cm particle board
between readout planes.
Active detector technologies being considered are resistive plate 
chambers and liquid and solid scintillator.  
\par The liquid scintillator design is for 14.4 m long multi-cell extrusions
of PVC, each containing 32 cells of width 3.75 cm.  The cells would be 3 cm
thick along the beam direction.  A looped fiber would be inserted in each cell
and an end-cap would be glued on one end, with a manifold/optical connector
assembly at the other.  There are no critical tolerances, such as positioning
of the fiber.  The proposal is under consideration at Fermilab.
\subsection{Brookhaven Wide Band Beam}
\par An alternative to the off-axis program has been developed at
Brookhaven.\cite{bib:bnl}  Using a wide-band beam, a huge detector such
as the UNO described in the last section of this paper, and a
longer distance to exploit possible matter effects, the Brookhaven idea
uses fits to the full energy spectrum to find and measure oscillation
parameters.  
\section{Reactor Experiments}
\par From the discovery of the neutrinos by Reines and Cowan\cite{bib:reines}
at Savannah River to the evidence for $\bar{\nu}_e$ disappearance at
KamLAND\cite{bib:kamland}, reactor neutrino experiments have studied
neutrinos in the same way -- observation of inverse beta decay with
scintillator detectors.  Since the signal from a reactor falls with
distance L as 1/L$^2$, as detectors have moved further away from
the reactors over the years, it has become more important to reduce
backgrounds.  That is achieved with a sufficient overburden,
and experiments one kilometer or more away from reactors (Chooz, Palo Verde
and KamLAND) have been
underground.
\par The KamLAND experiment measured a 40\% disappearance of
$\bar{\nu}_e$ presumably associated with the 2nd term in Equation \ref{eq:p}:
\begin{equation}
\label{eq:p}
P(\bar{\nu}_e \rightarrow \bar{\nu}_e) \cong 
- \sin^2 2 \theta_{13} \sin^2(\frac{\Delta m^2_{31} L}{4E})  - 
\cos^4 \quq \sin^2 2 \theta_{12} \sin^2(\frac{\Delta m^2_{21} L}{4E}) + 1
\end{equation}
The Chooz and Palo Verde data put a limit on $\quq$ (through the first
term in Equation \ref{eq:p}) of $\sin^2 2 \theta_{13} < 0.17$.  
Those experiments could not have had greatly improved sensitivity
to $\quq$ because of uncertainties related to knowledge of the flux of 
neutrinos from the reactors.  They were designed to test
whether the atmospheric neutrino anomaly might have been due to
$\nme$ oscillations, and hence were searching for large mixing.
\par Any new experiment to look for non-zero values of $\quq$ would
need the following properties:
\begin{itemize}
\item two or more detectors to reduce uncertainties to the reactor flux
\item identical detectors to reduce systematic errors 
\item carefully controlled energy calibration
\item low backgrounds and/or reactor-off data
\end{itemize}
\par
\begin{figure}
\vspace*{13pt}
\begin{center}
         \mbox{\epsfig{figure=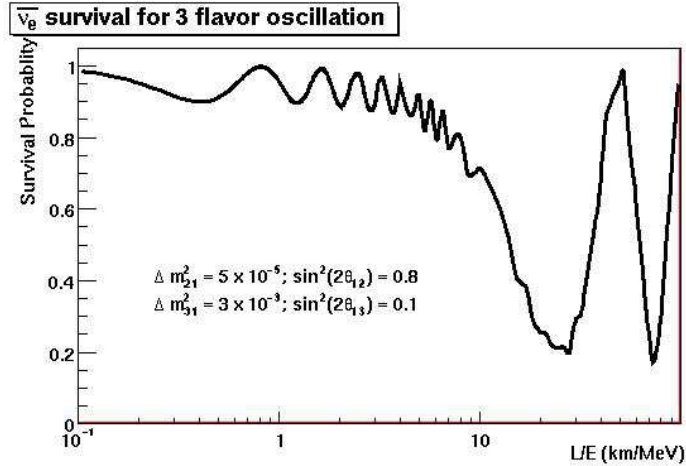,width=10.0cm}}
\caption{Probability of $\bar{\nu}_e$ disappearance versus L/E for 
$\quq$ near its current upper limit.}
\label{fig:P}
\end{center}
\end{figure}

 In Equation \ref{eq:p}, the values of $\theta_{12}$, $\Delta m^2_{21}$
and $\Delta m^2_{31}$ are approximately known.  In Figure \ref{fig:P},
The probability of $\bar{\nu}_e$ disappearance as a function of L/E
is plotted with  $\quq$ 
assumed to be near its maximum allowed value.  Note that
CP violation does not affect a disappearance experiment, and that
matter effects can be safely ignored in a reactor experiment.  The
large variation in P for L/E$>$ 10 km/MeV is the effect seen by
KamLAND and solar $\nu$ experiments.  The much smaller deviations from
unity for L/E $<$ 1 km/MeV are the goal for an accurate new reactor
experiment.

\par The optimization of detector distances for such a new experiment
is straightforward.  The statistical power comes from measuring a
deficit of $\bar{\nu}_e$ (up to a few percent) at the far detector,
along with a change in the energy spectrum consistent with that deficit.
Depending on the value of $\Delta m^2_{atm}$, the far detector should be
located about 1.7 km away.  For higher statistics experiments, the 
importance of the shape of the spectrum is greater than the rate, and
the optimum detector location moves about 30\% closer.
The optimum locations for detectors is thus
sensitive to eventual systematic errors as well as oscillation parameters.
But a near detector around 100m and a far detector around 1300 m will be
near the optimum.  Since the civil construction of laboratories 
might contribute half or more to the cost of an experiment, site
conditions could change the optimization.

\par A list of possible sites for a new reactor experiment is
included in Table \ref{tab:sites}, along with a tabulation of
previous reactor experiment sites.\cite{bib:white}  One obvious choice is 
the site of the 7 GW CHOOZ reactors in France, and a proposal called ``Double
CHOOZ" is being prepared.  A larger  15 ton far detector would be placed
in an existing hall 1050 m from the two reactors, and new near
detector would be placed close ($\sim 200m$).  The existing site
is an attraction, but is also a limitation.  An even larger detector
(50-1000 ton) would be desirable to achieve greater sensitivity on
$\quq$.  One of several projects being considered is to use the
7 GW Braidwood Reactors in Illinois, and place two shafts at 200 m and
the other at 1800 m, on average from the two reactor cores.  Two
25 ton detectors would be built.

\par 
A sensitivity of 0.02 in $\sin^2 2\quq$
can be achieved with as little as 250 ton-Gigawatt-years, while an
exposure of 8000 ton-Gigawatt-years may be required to achieve a
sensitivity of 0.01.\cite{bib:huber}
A two or more detector reactor experiment 
can find a non-zero
value for $\quq$ faster and less expensively than an off-axis experiment.
It does not face the degeneracies regarding CP parameters and the sign of
$\dmsq$, and hence cannot address those issues.  But a measurement of
$\quq$ by reactors followed by optimized off-axis experiments would together
measure neutrino parameters with much less uncertainty due to degeneracies
and correlations.

\begin{table}[htb]
\begin{center}
\begin{tabular}{|l|l|l|l|l|l|}\hline
Reactor & Location & L & Power & Overburden &  Mass \\ \hline \hline
        &          & m near/far & GW & MWE & ton \\ \hline
\multicolumn{6}{|c|}{ Previous Reactor Experiments} \\
Chooz & France & 1100  & 8.5  & 300  & 5  \\ \hline
Bugey & France & 49/95 & 5.6 & 16 & 1/0.5 \\ \hline
Palo Verde & Arizona & 890 & 11.6 & 32 & 11.3 \\ \hline
KamLAND & Japan & $<180>$ & 200 (26) & 2700 & 1000 \\ \hline \hline
\multicolumn{6}{|c|}{ Possible sites for New Reactor $\nu$ Experiments} \\
Angra & Brazil & 350/1350  & 4.0  & 60/600 & 50/50 \\ \hline
Braidwood & Illinois & 200/1800 & 7 & 250/250 & 25/50 \\ \hline
Double CHOOZ & France & 200/1050 & 7 & 50/300 & 10/10 \\ \hline
Daya Bay & China & 300/1500 & 11 & 200/1000 & 20/40 \\ \hline
Diablo Canyon & California & 400/1800 & 7 & 100/700 &25/50 \\ \hline
KASKA & Japan & 350/1300 & 24 & 140/600 & 8/8 \\ \hline
KR2DET & Russia & 115/1000 & 1.5 & 600/600 & 45/45 \\ \hline
\end{tabular}
\end{center}
\label{tab:sites}
\end{table}

\section{UNO nucleon decay project}
\par One of the reasons for the tremendous progress in understanding
the neutrino has been the fact that several detectors were built 
underground to search
for nucleon decay.  
The UNO detector\cite{bib:uno} is proposed as a next generation underground 
water Cerenkov detector that probes
nucleon decay beyond the sensitivities of the 
highly successful Super-Kamiokande  detector.\cite{bib:uno}  
The baseline 
conceptual design of the detector calls for a ``Multi-Cubical" design
with outer dimensions of 60x60x180 m$^3$. The 
detector has three optically independent cubical
compartments with corresponding photo-cathode coverage 
of 10\%, 40\%, and 10\%, respectively. 
The total (fiducial) mass of the detector is 650 (440) kton,
which is about 13 (20) times larger than 
the Super-K detector. 
Water Cerenkov technology is the only 
realistic detector technology available today to allow a
search for this decay mode for proton lifetimes 
up to 10$^{35}$ years. 
UNO provides other opportunities,
such as the ability to observe oscillatory behavior and   
appearance in the atmospheric neutrinos; precision measurement 
of temporal changes in the solar neutrino
fluxes; supernovae and supernova relic neutrinos;
and searches for astrophysical point sources 
of neutrinos.  UNO is an ideal distant detector for a long-baseline 
neutrino oscillation experiment with neutrino beam
energies below about 10 GeV, as envisaged by the Brookhaven working group.

\section{APS Neutrino Study}
A large number of ideas to study the neutrino sector are being pursued, and they
come with a variety of costs and feasibility.  Ideas for ``neutrino factories"
to study $\quq$ are now considered part of the program of the far future.  A
study in the United States to put the differing ideas for neutrino experiments
into a
single coherent program is being developed under the auspices of four divisions
of the American Physical Society.   Six working groups are holding
meetings and a report is due in the late summer of 2004.
Reactor experiments to see if $\quq$ is
non-zero, followed by accelerator projects to measure CP effects if it is,
should be a major part of such a program.

\end{document}